\documentclass[cameraready]{Interspeech}
\usepackage{cite}
\usepackage{makecell}
\title{SpiroPhonia: Non-Invasive Respiratory Health
Assessment from Spontaneous Speech}

\author[affiliation={1}, orcid=0009-0002-2605-5643, correspondingauthor]{Roksana}{Khanom}
\author[affiliation={2}, orcid=0009-0001-6053-9791]{Shafia}{Supty}
\author[affiliation={1},
orcid=0000-0001-5261-7780]{Nirupam}{Roy}
\author[affiliation={1}]{Ashok}{Agrawala}
\address{
    $^1$ University of Maryland, College Park, USA \\
    $^2$ DR. M R Khan Shishu Hospital \& Institute of Child Health, Bangladesh
}

\email{rkhanom@umd.edu, ssrd.supty@gmail.com, niruroy@umd.edu, agrawala@umd.edu }


\usepackage{comment}

\begin{document}

\maketitle

\begin{abstract}
    Chronic Obstructive Pulmonary Disease (COPD) remains a major global health challenge, emphasizing the need for accessible and non-invasive detection. Since speech production is fundamentally linked to respiratory physiology, its disruptions can serve as indirect indicators of pulmonary impairment. This study introduces SpiroPhonia, a machine learning framework that leverages spontaneous speech for respiratory health assessment. We evaluated SpiroPhonia on a new dataset of 201 speakers (102 with COPD, 99 healthy controls). By integrating statistical analysis with recursive feature selection, we identified a compact set of discriminative speech markers. Our best model achieved 78\% accuracy, 80\% F1-score, and 87\% AUC. This performance on spontaneous speech is competitive with methods using controlled laboratory recordings. Findings demonstrate that everyday speech encodes robust respiratory biomarkers, paving the way for continuous health monitoring via voice-enabled technologies.
\end{abstract}

\section{Introduction}
Chronic Obstructive Pulmonary Disease (COPD) is a progressive and largely irreversible respiratory disorder characterized by persistent respiratory symptoms and airflow limitation caused by structural abnormalities in the airways and alveoli \cite{GOLD2025,WHO2025}. COPD affects more than 400 million individuals worldwide and is the third leading cause of death globally \cite{DEOCA2025709}, degrading quality of life through chronic dyspnea, fatigue, and recurrent exacerbations \cite{chen2023global,10.1001/jamanetworkopen.2023.46598,wedzicha2007copd}. Early detection and proactive management can meaningfully slow disease progression and improve long-term outcomes \cite{lin2023current}. COPD diagnosis and monitoring remain fundamentally constrained by infrastructure-bound tools such as spirometry, which require clinical environments, trained personnel, and substantial patient cooperation. This creates a critical sensing and accessibility gap: early-stage disease often remains undiagnosed\cite{lamprecht2011copd}, and longitudinal monitoring is episodic rather than continuous. These limitations motivate the need for scalable, passive, and at-home respiratory assessment modalities\cite{al2016monitoring}, including speech-based and acoustic biomarkers, that can enable unobtrusive screening and continuous monitoring outside traditional clinical settings.

Speech production is tightly coupled with respiratory physiology: airflow from the lungs drives phonation, and subglottal pressure regulates vocal stability and timing \cite{hansen2018focus,kent2004uniqueness,fant1971acoustic,maryn2010toward,elshebl2025study}. In individuals with COPD, airflow obstruction alters phonatory excitation and breath–speech coordination, producing measurable changes in pitch variability, spectral energy distribution, and pause behavior. This strong physiological coupling provides an opportunity to detect respiratory impairment through acoustic and prosodic patterns embedded in natural speech\cite{zeng2023exploring,farrus2021speech,mayr2025assessing}.
\begin{figure}[!t]
    \centering
    \includegraphics[width=0.95\columnwidth]{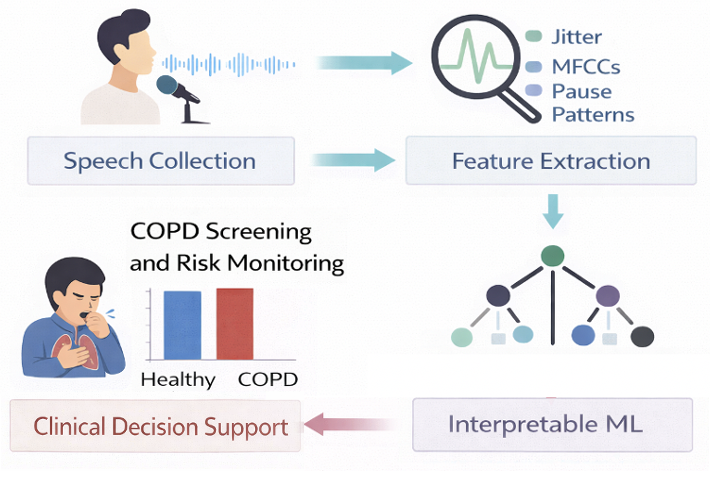}
    \vspace{-0.15in}
    \caption{SpiroPhonia framework for respiratory health assessment from spontaneous speech.}
    \vspace{-0.1in}
    \label{fig:hero}
    \vspace{-0.15in}
\end{figure}
Prior research has explored the potential of speech-based respiratory monitoring as an accessible alternative to clinic-based assessments. Nathan et al.\cite{nathan2019assessment} investigated pulmonary disease assessment using biomarkers extracted from scripted natural speech in a cohort of 131 participants. Similarly, Chun et al.\cite{san2020towards} examined passive pulmonary function using scripted speech collected in controlled clinical environments. In parallel, Idrisoglu et al.\cite{idrisoglu2025vowel} and Triantafyllopoulos et al.\cite{triantafyllopoulos2024sustained} demonstrated COPD classification using sustained phonation tasks. Although these studies demonstrate the feasibility of acoustic biomarkers, existing approaches primarily rely on elicited speech tasks, such as sustained vowels, scripted reading, or guided breathing maneuvers, recorded in structured or laboratory settings\cite{sankey2025detecting,saleheen2020lung,xu2021listen2cough}. These protocols require explicit user participation and adherence to prescribed procedures, which may limit their suitability for continuous, at-home monitoring and long-term respiratory surveillance.

In contrast, spontaneous conversational speech is naturally produced during daily interactions and requires no structured tasks or controlled environments. The widespread availability of microphones embedded in smartphones and voice-enabled systems (e.g., Alexa, Siri) ensures the continuous generation of natural speech in everyday contexts, even under noisy conditions\cite{shahid2022voicefind}. Prior work has also demonstrated that natural sound carries rich information about subtle physical states of the source and physical environment beyond spoken content, enabling robust inference of physical interactions, environmental context, and user activities from everyday acoustic signals \cite{takawale2025infer,shahid2025human,mishrasing,bai2025sparse,bai2024scribe,takawale2024learning,shahid2023my,bai2023natural,garg2023structure,bai2022spidr,garg2021owlet,sami2020spying,garg2020acoustic,shen2018mute,roy2017backdoor,roy2016ripple,roy2015ripple}. These findings suggest that spontaneous speech may similarly encode latent biomarkers associated with respiratory health.
If spontaneous speech encodes stable and clinically meaningful respiratory biomarkers, existing voice-enabled infrastructure could support real-time, at-home monitoring by tracking longitudinal changes in vocal biomarkers without additional hardware or behavioral requirements. Nevertheless, spontaneous speech exhibits variability in content, prosody, speaker traits, and recording conditions, which may obscure subtle respiratory signals and challenge subject-independent modeling. This raises a fundamental question: whether real-world conversational speech contains robust biomarkers that remain discriminative under uncontrolled conditions.

In this work, we address this challenge by investigating spontaneous speech as a substrate for COPD detection under rigorous subject-independent evaluation. We curate a dataset of 201 speakers (102 with COPD or related respiratory conditions and 99 healthy controls) derived from real-world recordings, with respiratory status labels verified by a respiratory physician. We identify and validate a compact set of acoustic, spectral, and temporal biomarkers that remain robust in uncontrolled settings, reducing reliance on structured speech tasks and specialized equipment. Furthermore, we develop an interpretable machine learning framework (Figure 1) that achieves strong performance (78\% accuracy, 80\% F1-score, 87\% AUC) and outline a practical pathway toward deployment through screening-oriented and on-device model configurations, a design that deliberately favors physiologically grounded predictors over opaque deep data-driven alternatives.

\section{Methods}
\subsection{SpiroPhonia Dataset Curation and Collection}
Most prior work on speech-based respiratory health monitoring has relied on data collected in clinical or laboratory environments, which are expensive to run and often not publicly shared due to privacy and institutional constraints. This limits access to large scale respiratory speech corpora and hinders replication and secondary analysis. To address these challenges, we constructed the SpiroPhonia dataset using publicly available spontaneous speech sourced from online platforms, which host a substantial volume of unscripted, real-world speech. As of mid-2025, social media platforms reach over 5.4 billion users, and YouTube alone receives more than 500 hours of video uploads per minute\cite{DataReportal2025,OberloYouTubeStats2025}. Prior research has successfully used such content for health analysis, including COVID-19 detection~\cite{triantafyllopoulos2022covyt,anibal2024omicron}, autism assessment~\cite{fusaro2014potential}, and mental health monitoring \cite{abhishek2021depiction,zhang2020relationships}.

Motivated by these findings, we focus on spontaneous conversational speech in English as a natural and abundant source of information for respiratory health analysis. The SpiroPhonia dataset was assembled by identifying interviews and personal narratives in which speakers self identified as having COPD, asthma, emphysema, or other chronic respiratory conditions. Targeted keyword queries such as “COPD patient interview” and “lung disease story” were used to locate candidate recordings. 
Healthy control samples were drawn from interviews with individuals of a comparable age range to the patient cohort, including both speakers in professions requiring sustained respiratory fitness and individuals who explicitly reported no known respiratory conditions. All video recordings were manually inspected for audio quality and conversational content, and respiratory status labels were subsequently verified by a respiratory physician based on contextual content. Only high quality, publicly available conversational speech was included. As all recordings are publicly available and analyzed retrospectively without interaction or re-identification, this study falls under exempt human-subjects research and does not require informed consent or institutional review board approval. No personally identifiable information was retained. The final dataset comprises 201 speakers, including 102 individuals with respiratory conditions and 99 healthy controls. Table~\ref{tab:dataset_summary} summarizes the dataset composition. \textbf{Dataset availability:} Available upon request for academic research purposes.

\begin{table}[htbp]
\centering
\vspace{-0.05in}
\caption{Summary of the SpiroPhonia dataset.}
\vspace{-0.1in}
\label{tab:dataset_summary}
\renewcommand{\arraystretch}{1.15}
\resizebox{\columnwidth}{!}{%
\begin{tabular}{l c c}
\hline
\textbf{Category} & \textbf{COPD/Respiratory Patients} & \textbf{Healthy Controls} \\
\hline
Number of Subjects & 102 & 99 \\
Age Range (years) & \multicolumn{2}{c}{45--95} \\
Language & \multicolumn{2}{c}{English} \\
Male (n) & 50 & 47 \\
Female (n) & 52 & 52 \\
\hline
\end{tabular}%
}
\vspace{-0.2in}
\end{table}
%\hline
\subsection{Preprocessing}
\begin{figure}[t]
\centering
\includegraphics[width=\columnwidth]{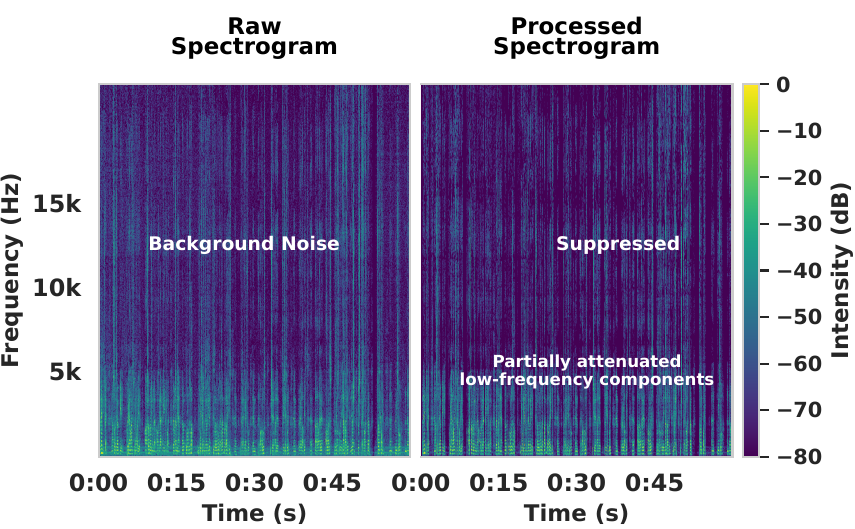}
\vspace{-0.2in}
\caption{Spectrograms of a 70-year-old male’s spontaneous speech before and after signal processing. }
\label{fig:spectrogram}
\vspace{-0.2in}
\end{figure}

All recordings were converted to 44.1 kHz, 16-bit mono WAV format. Non-speech content (interviewer prompts, long silences, overlapping background voices) was removed using energy-based voice activity detection followed by manual refinement. Clean speech was divided into 10–30 s windows: shorter segments yield unstable pause-based estimates due to the sparsity of pause events, while longer segments introduce topic- and style-driven prosodic variability unrelated to respiratory state. This range balances feature stability with natural prosodic variation. Each segment was amplitude-normalized, band-pass filtered between 100 and 5000 Hz, and denoised with light spectral subtraction when necessary (Figure~\ref{fig:spectrogram}), ensuring reliable extraction of acoustic, spectral, and temporal features.

%\begin{comment}
%\end{comment}
\subsection{Speech Feature Extraction and Analysis}
We extracted three complementary groups of features from each preprocessed segment: acoustic perturbation features, spectral representations, and pulmonary-inspired temporal features. Together they capture vocal fold stability, vocal tract shaping, and breath-speech coordination. 

\textbf{Acoustic features:} We extracted acoustic features that quantify pitch stability, amplitude regularity, and voice quality. We estimated the fundamental frequency ($F_{0}$) as the inverse of the average glottal period within voiced regions and computed its mean and variance.
\begin{table}[t]
\centering
\caption{Statistical analysis of key acoustic, spectral, and temporal features between COPD and healthy speakers.}
\label{tab:key_stats}
\renewcommand{\arraystretch}{1.1}
\resizebox{\columnwidth}{!}{%
\begin{tabular}{p{1.8cm} p{2.4cm} p{4.0cm} p{1.3cm} c c}
\hline
\textbf{Group} & \textbf{Feature} & \textbf{Description} & \textbf{p-value} & \textbf{Sig.} & \textbf{Effect size} \\
\hline
Acoustic &
F0 (Mean) &
Average pitch (Hz) &
0.0032 & ** & 0.28 \\

Acoustic &
RAP Jitter &
Cycle to cycle frequency perturbation &
$<$0.001 & *** & 0.36 \\

Acoustic &
DDP Jitter &
Three cycle RAP jitter &
$<$0.001 & *** & 0.36 \\

Acoustic &
PPQ5 Jitter &
Five cycle pitch perturbation quotient &
$<$0.001 & *** & 0.35 \\

Acoustic &
Shimmer (Local) &
Amplitude perturbation &
0.046 & * & 0.21 \\

Spectral (MFCC) &
MFCC10 (Mean) &
Mean mid frequency cepstral energy &
$<$0.001 & *** & 0.62 \\

Spectral (MFCC) &
MFCC10 (Min) &
Minimum mid frequency cepstrum &
$<$0.001 & *** & 0.61 \\

Spectral (MFCC) &
MFCC10 (Median) &
Median mid frequency cepstrum &
$<$0.001 & *** & 0.59 \\

Spectral (MFCC) &
MFCC3 (Mean) &
Mean low order cepstral component &
$<$0.001 & *** & 0.59 \\

Spectral (MFCC) &
MFCC3 (SD) &
Variation of low order cepstrum &
0.0003 & ** & 0.53 \\

Spectral (MFCC) &
MFCC6 (Min) &
Minimum mid low cepstral band &
0.0014 & ** & 0.46 \\

Spectral (MFCC) &
MFCC7 (Min) &
Minimum mid frequency band &
0.0001 & *** & 0.57 \\

Spectral (MFCC) &
MFCC7 (Mean) &
Mean mid frequency cepstrum &
0.0002 & ** & 0.55 \\

Spectral (MFCC) &
MFCC11 (SD) &
High frequency cepstral variation &
$<$0.001 & *** & 0.37 \\

Spectral (MFCC) &
MFCC13 (SD) &
High order cepstral variability &
0.0003 & ** & 0.52 \\

Temporal &
Pause Ratio &
Fraction of total speech time in silence &
$<$0.001 & *** & 0.43 \\

Temporal &
Pauses per Minute &
Number of pauses per minute of speech &
$<$0.001 & *** & 0.61 \\

Temporal &
Total Pause Duration &
Total duration of all silence segments (s) &
0.002 & ** & 0.40 \\
\hline
\end{tabular}%
}
\vspace{1pt}
{\footnotesize *p$<$0.05, **p$<$0.01, ***p$<$0.001. Effect sizes reported as Cohen's d (parametric) or rank biserial correlation (non parametric).}
\vspace{-0.2in}
\end{table}
We quantified cycle-to-cycle frequency and amplitude perturbations using standard jitter measures (local jitter, RAP, and PPQ5), shimmer, and the harmonic-to-noise ratio (HNR)\cite{yumoto1982harmonics}. We also extracted formant frequencies ($F_{1}$--$F_{4}$) to characterize vocal-tract resonances\cite{titze1993vocal,gauffin1989spectral}.

\textbf{Spectral features:} Spectral structure was characterized using Mel Frequency Cepstral Coefficients (MFCCs) \cite{rabiner2010theory}.
%\cite{davis1980comparison,rabiner2010theory,eyben2010opensmile}
For each segment, 13 MFCCs were computed, and five summary statistics were extracted per coefficient, yielding 65 spectral features capturing vocal-tract resonance and fine-grained spectral variation \cite{tzirakis2017end}.

\textbf{Temporal features:} To capture respiration-related timing structure, we extracted pause-based features using an energy threshold of $-40$\,dB and a minimum pause duration of $100$\,ms \cite{rabiner1978digital,huber2008effects}. From the detected pauses, we computed pause ratio (fraction of total duration occupied by silence), pauses per minute, and total pause duration. These measures have been associated with respiratory effort and airflow limitation in COPD \cite{werner2023phonetics,mayr2025assessing}.
\subsection{Statistical Analysis and Summary of Feature Behavior}

We evaluated the discriminative power of each feature using a two-stage statistical analysis, as shown in Table~\ref{tab:key_stats}. Normality was assessed with the Shapiro--Wilk test \cite{razali2011power,shapiro1965analysis}, followed by either independent-samples $t$-tests or Mann--Whitney U tests depending on distribution characteristics \cite{gibbons1992nonparametric,mann1947test}. Bonferroni correction was applied with a significance threshold of $p<0.05$\cite{bland1995multiple}. Effect sizes were reported as Cohen's $d$ for parametric tests and rank-biserial correlation for non-parametric tests \cite{cohen2013statistical,kerby2014simple}. 

\begin{figure}[t]
\centering
\includegraphics[width=1\columnwidth,height=\textheight,
  keepaspectratio]{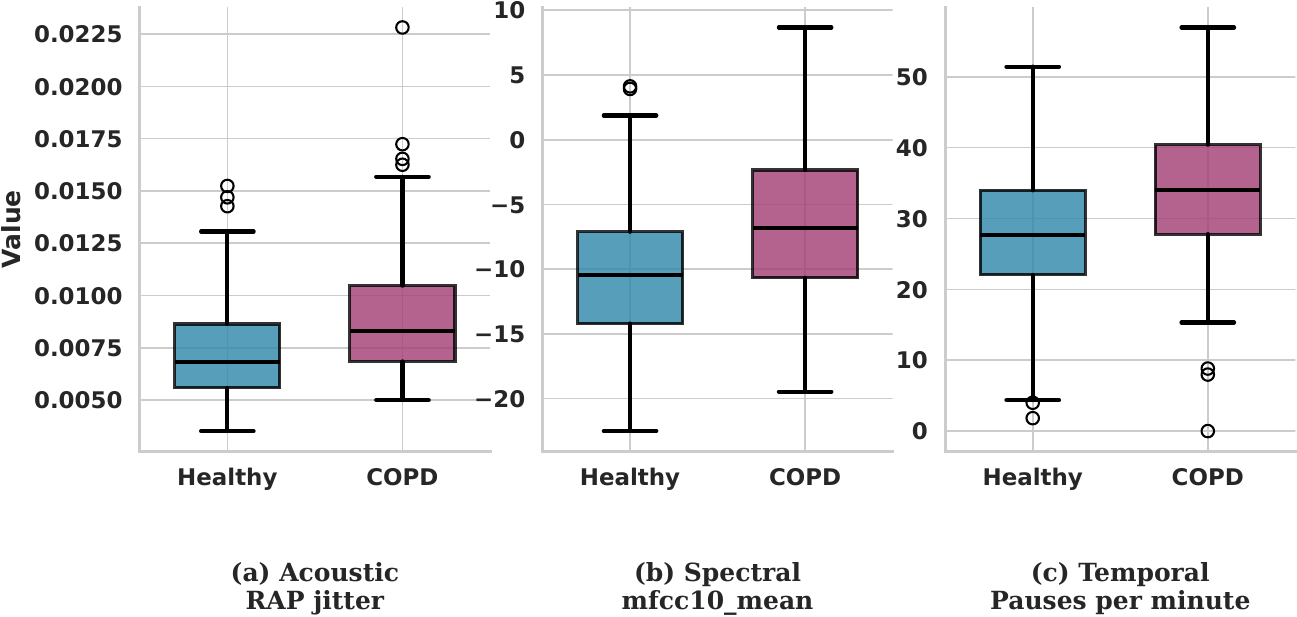}
\vspace{-0.2in}
\caption{Distribution of key acoustic, spectral, and temporal speech features for COPD and healthy speakers.}
\vspace{-0.07in}
\label{fig:feature-families}
\end{figure}

Across 25 acoustic features, jitter-based parameters showed the strongest between-group differences ($p<0.001$), with small-to-moderate effect sizes. Mean $F_{0}$ was slightly lower for COPD speakers, while shimmer showed a consistent but weaker elevation. Among the 65 MFCC-derived spectral features, several mid- and high-order coefficients (notably MFCC10 and MFCC3) exhibited robust separation with effect sizes up to $d \approx 0.6$. All three temporal pause features were significantly different, with pause ratio and pauses per minute showing the highest rank-biserial correlations, indicating reduced respiratory efficiency and irregular breath--speech coordination in COPD speech. Figure \ref{fig:feature-families} shows representative features with distributional shifts between groups.

\begin{table}[t]
\centering
\caption{Classification performance and bootstrap 95\% confidence intervals.}
\vspace{-0.1in}
\label{tab:performance}
\renewcommand{\arraystretch}{1.1}
\setlength{\tabcolsep}{3pt}
\resizebox{\columnwidth}{!}{%
\begin{tabular}{l c c c c c c c}
\hline
\textbf{Classifier} & \textbf{Feat.} & \textbf{Acc.} & \textbf{Bal. Acc.} & \textbf{AUC} & \textbf{Sens.} & \textbf{Spec.} & \textbf{F1} \\
 & & \scriptsize{[95\% CI]} & \scriptsize{[95\% CI]} & \scriptsize{[95\% CI]} & \scriptsize{[95\% CI]} & \scriptsize{[95\% CI]} & \scriptsize{[95\% CI]} \\
\hline
Linear SVM & 5
& \makecell{\textbf{78.05}\\\scriptsize{[65.8, 90.2]}}
& \makecell{\textbf{78.21}\\\scriptsize{[64.5, 90.1]}}
& \makecell{79.05\\\scriptsize{[63.8, 92.2]}}
& \makecell{71.43\\\scriptsize{[52.2, 89.5]}}
& \makecell{\textbf{85.00}\\\scriptsize{[66.6, 100]}}
& \makecell{76.92\\\scriptsize{[58.8, 89.8]}} \\
Gradient Boosting & 59
& \makecell{\textbf{78.05}\\\scriptsize{[65.9, 90.2]}}
& \makecell{77.86\\\scriptsize{[65.3, 90.1]}}
& \makecell{84.76\\\scriptsize{[71.3, 96.7]}}
& \makecell{\textbf{85.71}\\\scriptsize{[68.7, 100]}}
& \makecell{70.00\\\scriptsize{[50.0, 88.2]}}
& \makecell{\textbf{80.00}\\\scriptsize{[65.1, 92.0]}} \\
Random Forest & 59
& \makecell{75.61\\\scriptsize{[63.4, 87.8]}}
& \makecell{75.36\\\scriptsize{[62.1, 87.9]}}
& \makecell{\textbf{87.14}\\\scriptsize{[74.6, 96.8]}}
& \makecell{\textbf{85.71}\\\scriptsize{[68.4, 100]}}
& \makecell{65.00\\\scriptsize{[42.9, 86.7]}}
& \makecell{78.26\\\scriptsize{[63.4, 89.8]}} \\
Logistic Regression & 14
& \makecell{73.17\\\scriptsize{[58.5, 85.4]}}
& \makecell{73.10\\\scriptsize{[58.8, 85.8]}}
& \makecell{80.48\\\scriptsize{[65.7, 92.4]}}
& \makecell{76.19\\\scriptsize{[56.3, 94.1]}}
& \makecell{70.00\\\scriptsize{[47.8, 89.5]}}
& \makecell{74.42\\\scriptsize{[57.1, 87.5]}} \\
Neural Network$^{*}$ & 14 & 70.73 & 70.83 & 78.10 & 66.67 & 75.00 & 70.00 \\
\hline
\end{tabular}%
}
\vspace{-0.05in}\\
\footnotesize{$^{*}$Point estimates only due to training stochasticity.}
\vspace{-0.2in}
\end{table}

\section{Results}
\subsection{Model Design and Feature Optimization}
We employed a two-stage feature optimization strategy combining statistical refinement with model-based selection. The dataset was first partitioned into a 160-sample training set and a 41-sample held-out test set via a stratified split, ensuring strict subject-independent evaluation. All feature processing and model fitting were confined to the training set; the held-out samples were used solely for final evaluation. In the first stage, the 94 acoustic, spectral, and temporal attributes were filtered by removing near-zero variance features and pruning correlated predictors\cite{kuhn2013applied}. For pairs with absolute Pearson correlation greater than 0.9, we retained the feature with the larger effect size from the statistical analysis in Section 2.4, yielding 67 non-redundant features. In the second stage, we applied Recursive Feature Elimination with Cross-Validation (RFECV)\cite{guyon2002gene,pedregosa2011scikit} using a stratified five-fold scheme, with ranking and fitting restricted to each fold's training portion to prevent leakage\cite{varma2006bias,cawley2010over}. The optimal subset size was selected by maximizing mean cross-validation accuracy. As the recursive ranking criterion is estimator-dependent, RFECV was applied independently per classifier, yielding classifier-specific subsets (Table~\ref{tab:features}). Default scikit-learn configurations were used; feature subset size was the only tuned hyperparameter.

\begin{figure}[ht]
\centering
\vspace{-0.2pt}
\includegraphics[
  width=0.9\columnwidth
  %height=0.5\textheight
  %keepaspectratio
]{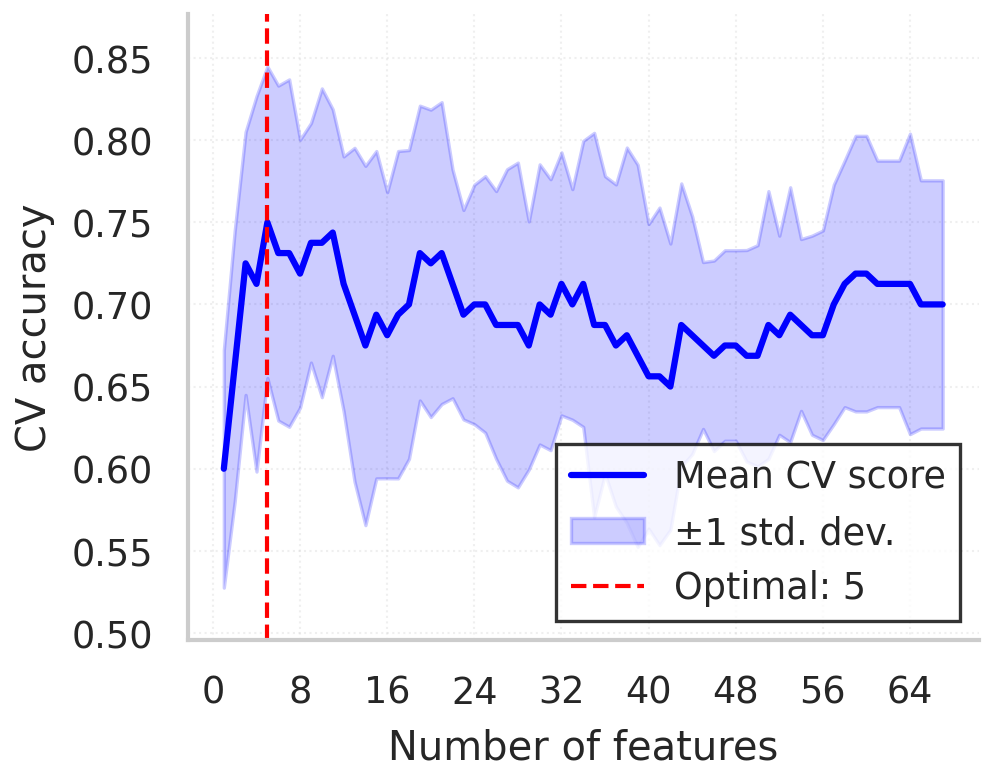}
\vspace{-0.15in}
\caption{RFECV performance curve showing cross-validated accuracy versus number of selected features for the linear SVM.}
\label{fig:rfecv_curve}
\vspace{-0.2in}
\end{figure}
%\vspace{-0.2pt}

\subsection{Classification Performance} 
We evaluated five machine learning classifiers for COPD detection using the SpiroPhonia dataset. All models were trained on the 160-sample training set and evaluated on the 41-sample held-out test set described in Section 3.1. Performance metrics and bootstrap 95\% confidence intervals (1{,}000 speaker-level resamples) are summarized in Table~\ref{tab:performance}. The resulting intervals were consistent with the classifier ranking observed from the point estimates. Linear Support Vector Machine (SVM) and Gradient Boosting achieved the highest accuracy (78.05\%) under distinct operating profiles, with high specificity for SVM (85.00\%) using only five features and high sensitivity for Gradient Boosting (85.71\%, 95\% CI: 68.74--100.00\%), while Random Forest achieved the highest AUC (87.14\%, 95\% CI: 74.61--96.77\%). Notably, all classifiers maintained accuracy lower bounds above 58\%, and the top-performing models retained AUC lower bounds above 71\%, indicating that the observed performance reflects a stable and meaningful discriminative signal rather than an artifact of a particular test-set composition.

\subsection{Discriminative Speech Biomarkers} %Discriminative Feature Patterns and Key Predictive Markers%}

The feature optimization process revealed that most discriminative information is concentrated within a compact subset of predictors. As shown in the RFECV curve (Figure~\ref{fig:rfecv_curve}), cross validated accuracy increases sharply with the first few features and stabilizes beyond approximately ten, indicating that only a small number of acoustic and spectral cues are required for reliable COPD detection. Linear models, especially the SVM, achieved strong performance with only 5 features, highlighting an advantageous balance between parsimony and discriminative capability. Logistic Regression and the shallow Neural Network converged around 14 features, while ensemble methods such as Random Forest and Gradient Boosting required substantially larger subsets, retaining roughly 59 predictors to capture the nonlinear structure present in spontaneous speech.

Despite differences in model complexity, several predictors consistently emerged as robust markers of respiratory impairment across classifiers. Local jitter, local shimmer, and pause ratio were selected by every model, reflecting their sensitivity to physiological changes associated with COPD. 
A group of mid order MFCCs (including MFCC2-MFCC5 and MFCC10-MFCC12) also appeared frequently across models, indicating that spectral shaping and turbulence related energy patterns provide complementary information about airflow limitation.
Table~\ref{tab:features} summarizes the feature subsets selected by each classifier.

\begin{table}[t]
\centering
\caption{Feature selection patterns across classifiers.}
\vspace{-0.1in}
\label{tab:features}
\renewcommand{\arraystretch}{1.2}
\resizebox{\columnwidth}{!}{%
\begin{tabular}{l c p{5cm}}
\hline
\textbf{Classifier} & \textbf{Features} & \textbf{Key Features and Performance Characteristics} \\
\hline
Linear SVM & 5 & Local Jitter, Local Shimmer, Pause Ratio, MFCC2 Std, ShimmerPCA. Highest specificity and feature efficiency. \\
Gradient Boosting & 59 & Comprehensive acoustic \& spectral set. Best F1 score and sensitivity. \\
Random Forest & 59 & F0, HNR, Jitter and Shimmer, Formants. Highest AUC and strong sensitivity. \\
Logistic Regression & 14 & Local Jitter, Local Shimmer, F3 Mean, Pause Ratio, MFCC features. \\
Neural Network & 14 & Similar to Logistic Regression with MFCC emphasis. Moderate performance, limited by dataset size. \\
\hline
\vspace{-0.4in}
\end{tabular}%
}
\end{table}
\subsection{Clinical Utility Assessment}
The clinical relevance of each classifier was examined with respect to its operating characteristics and potential deployment context. Distinct patterns of performance indicate that different models may serve complementary roles within a respiratory-health monitoring framework. Gradient Boosting, with its high sensitivity, is suitable for initial screening where minimizing false negatives is critical. Random Forest, achieving the highest AUC, is more appropriate for high-confidence detection scenarios requiring robust class separation. Linear SVM, combining high specificity with only five features, is preferable for confirmatory assessments and resource-constrained mobile deployment. This tiered interpretation underscores that optimal model selection depends on the intended clinical application, enabling adaptive deployment across screening, diagnostic, and follow-up stages of care.

\subsection{Comparative Evaluation}

Prior studies report COPD detection accuracies between 67\% and 84.6\% using sustained vowels or scripted speech collected under controlled laboratory or clinical conditions. In contrast, as shown in Table~\ref{tab:comparison}, our approach operates on fully spontaneous speech recorded in uncontrolled real-world environments. Despite this increased variability, the proposed compact feature sets achieve performance comparable to controlled-setting approaches, attaining 78.05\% accuracy, demonstrating that these features remain robust and discriminative for real-world respiratory assessment, with reported confidence intervals reflecting typical small-cohort variance in this domain.

\begin{table}[!t]
\centering
\vspace{0.2in}
\caption{Performance comparison across prior studies.}
\label{tab:comparison}
\vspace{-0.1in}
\renewcommand{\arraystretch}{1.15}
\resizebox{\columnwidth}{!}{%
\begin{tabular}{p{3.5cm} p{2.5cm} p{1.7cm} p{3.2cm}}
\hline
\textbf{Study} & \textbf{Speech Type} & \textbf{Accuracy} & \textbf{Data Collection} \\
\hline
\textbf{Our Study} & \textbf{Spontaneous} & \textbf{78.05\%} & \textbf{Real-world interviews} \\
Idrisoglu et al.\cite{idrisoglu2025vowel} & Vowel segments & 84.6\% & Laboratory controlled \\
%Triantafyllopoulos et al. \cite{triantafyllopoulos2024sustained} & Sustained vowels & 79.0\% & Clinical setting \\
Sankey-Olsen et al.\cite{sankey2025detecting} & Read speech & 67.0\% & Controlled environment \\
Chun et al.\cite{san2020towards} & Spontaneous & 75.2\% & clinical setting\\
\hline
\end{tabular}
}
\vspace{-0.2in}
\end{table}

\section{Conclusion}
This paper demonstrates that spontaneous conversational speech contains consistent acoustic and prosodic signatures of respiratory impairment, enabling subject-independent discrimination between COPD and healthy speakers under real-world recording variability. The proposed two-stage feature optimization framework identified a compact and interpretable set of predictors reflecting phonatory instability, altered spectral structure, and disrupted breath speech coordination. Competitive performance with minimal features supports the feasibility of efficient respiratory assessment from everyday speech. The current limitations and opportunities for improvement include the modest dataset size, which restricts the evaluation of more data-intensive modeling approaches, and the need for broader validation across languages, recording devices, and acoustic environments to assess robustness and generalizability at scale.
Our future work will focus on dataset expansion and systematic cross-environment evaluation to strengthen clinical reliability.

\section{Acknowledgments}

This work was partially supported by the NSF CAREER Award 2238433. We also thank the various companies that sponsor the MIND Lab and the iCoSMoS Lab at UMD.

\section{Generative AI Use Disclosure}
Generative AI tools were used solely for language editing and polishing. All scientific content, experimental design, data analysis, and interpretations were conceived and performed by the authors. The authors take full responsibility for its content.
%of this paper.

\bibliographystyle{IEEEtran}
\bibliography{IEEEfull.bib,ref_acoustics_Nirupam_Jun2026}

\end{document}